\documentclass[epj]{svjour}

\usepackage{graphicx}

\usepackage{amsmath}
\usepackage{bm}
\usepackage{ifthen} % for conditional statements
\newboolean{uprightparticles}
\setboolean{uprightparticles}{false} %True for upright particle symbols
\usepackage{xspace} 
\usepackage{upgreek}

\def\lhcb   {\mbox{LHCb}\xspace}

\def\babar  {\mbox{BaBar}\xspace}
\def\belle  {\mbox{Belle}\xspace}
\def\belletwo {\mbox{Belle~II}\xspace}

\def\MagUp {\mbox{\em Mag\kern -0.05em Up}\xspace}

\ifthenelse{\boolean{uprightparticles}}%
{

 \def\Ppi         {\ensuremath{\uppi}\xspace}                 
                  
 \def\Prho        {\ensuremath{\uprho}\xspace}

 \def\Ppsi        {\ensuremath{\uppsi}\xspace}

 \def\PDelta      {\ensuremath{\Delta}\xspace}                 
 \def\PXi         {\ensuremath{\Xi}\xspace}                 
 \def\PLambda     {\ensuremath{\Lambda}\xspace}                 
 \def\PSigma      {\ensuremath{\Sigma}\xspace}                 
 \def\POmega      {\ensuremath{\Omega}\xspace}                 
 \def\PUpsilon    {\ensuremath{\Upsilon}\xspace}

 \def\PB      {\ensuremath{\mathrm{B}}\xspace}                 
                  
 \def\PD      {\ensuremath{\mathrm{D}}\xspace}

 \def\PJ      {\ensuremath{\mathrm{J}}\xspace}                 
 \def\PK      {\ensuremath{\mathrm{K}}\xspace}

 \def\Pb      {\ensuremath{\mathrm{b}}\xspace}                 
                  
 \def\Pd      {\ensuremath{\mathrm{d}}\xspace}                 
 \def\Pe      {\ensuremath{\mathrm{e}}\xspace}

 \def\Pi      {\ensuremath{\mathrm{i}}\xspace}

 \def\Ps      {\ensuremath{\mathrm{s}}\xspace}                 
 \def\Pt      {\ensuremath{\mathrm{t}}\xspace}                 
 \def\Pu      {\ensuremath{\mathrm{u}}\xspace}

 \def\thebaroffset{0.0em}
}
{

 \def\Ppi         {\ensuremath{\pi}\xspace}                 
                  
 \def\Prho        {\ensuremath{\rho}\xspace}

 \def\Ppsi        {\ensuremath{\psi}\xspace}                 
                  
 \mathchardef\PDelta="7101
 \mathchardef\PXi="7104
 \mathchardef\PLambda="7103
 \mathchardef\PSigma="7106
 \mathchardef\POmega="710A
 \mathchardef\PUpsilon="7107
                  
 \def\PB      {\ensuremath{B}\xspace}                 
                  
 \def\PD      {\ensuremath{D}\xspace}

 \def\PJ      {\ensuremath{J}\xspace}                 
 \def\PK      {\ensuremath{K}\xspace}

 \def\Pb      {\ensuremath{b}\xspace}                 
                  
 \def\Pd      {\ensuremath{d}\xspace}                 
 \def\Pe      {\ensuremath{e}\xspace}

 \def\Pi      {\ensuremath{i}\xspace}

 \def\Ps      {\ensuremath{s}\xspace}                 
 \def\Pt      {\ensuremath{t}\xspace}                 
 \def\Pu      {\ensuremath{u}\xspace}

 \def\thebaroffset{0.18em}
}
\newcommand{\offsetoverline}[2][\thebaroffset]{\kern #1\overline{\kern -#1 #2}}%

\makeatletter
\ifcase \@ptsize \relax% 10pt
  \newcommand{\miniscule}{\@setfontsize\miniscule{4}{5}}% \tiny: 5/6
\or% 11pt
  \newcommand{\miniscule}{\@setfontsize\miniscule{5}{6}}% \tiny: 6/7
\or% 12pt
  \newcommand{\miniscule}{\@setfontsize\miniscule{5}{6}}% \tiny: 6/7
\fi
\makeatother

\DeclareRobustCommand{\optbar}[1]{\shortstack{{\miniscule (\rule[.5ex]{1.25em}{.18mm})}
  \\ [-.7ex] $#1$}}

\def\epem       {{\ensuremath{\Pe^+\Pe^-}}\xspace}
\def\uquark    {{\ensuremath{\Pu}}\xspace}
\def\uquarkbar {{\ensuremath{\overline \uquark}}\xspace}

\def\dquark    {{\ensuremath{\Pd}}\xspace}
\def\dquarkbar {{\ensuremath{\overline \dquark}}\xspace}

\def\squark    {{\ensuremath{\Ps}}\xspace}

\def\bquark    {{\ensuremath{\Pb}}\xspace}
\def\bquarkbar {{\ensuremath{\overline \bquark}}\xspace}

\def\tquark    {{\ensuremath{\Pt}}\xspace}

\def\pion   {{\ensuremath{\Ppi}}\xspace}
\def\piz    {{\ensuremath{\pion^0}}\xspace}
\def\pip    {{\ensuremath{\pion^+}}\xspace}
\def\pim    {{\ensuremath{\pion^-}}\xspace}

\def\pimp   {{\ensuremath{\pion^\mp}}\xspace}

\def\rhomeson {{\ensuremath{\Prho}}\xspace}
\def\rhoz     {{\ensuremath{\rhomeson^0}}\xspace}
\def\rhop     {{\ensuremath{\rhomeson^+}}\xspace}
\def\rhom     {{\ensuremath{\rhomeson^-}}\xspace}

\def\kaon    {{\ensuremath{\PK}}\xspace}
\def\KorKbar {\kern \thebaroffset\optbar{\kern -\thebaroffset \PK}{}\xspace}

\def\KS      {{\ensuremath{\kaon^0_{\mathrm{S}}}}\xspace}

\def\D       {{\ensuremath{\PD}}\xspace}

\def\DorDbar {\kern \thebaroffset\optbar{\kern -\thebaroffset \PD}\xspace}

\def\Dp      {{\ensuremath{\D^+}}\xspace}
\def\Dm      {{\ensuremath{\D^-}}\xspace}

\def\DpDm    {\ensuremath{\Dp {\kern -0.16em \Dm}}\xspace}

\def\B       {{\ensuremath{\PB}}\xspace}
\def\Bbar    {{\ensuremath{\offsetoverline{\PB}}}\xspace}

\def\BorBbar {\kern \thebaroffset\optbar{\kern -\thebaroffset \PB}\xspace}
\def\Bz      {{\ensuremath{\B^0}}\xspace}
\def\Bzb     {{\ensuremath{\Bbar{}^0}}\xspace}
\def\Bd      {{\ensuremath{\B^0}}\xspace}

\def\BdorBdbar {\kern \thebaroffset\optbar{\kern -\thebaroffset \Bd}\xspace}
\def\Bu      {{\ensuremath{\B^+}}\xspace}
\def\Bub     {{\ensuremath{\B^-}}\xspace}
\def\Bp      {{\ensuremath{\Bu}}\xspace}
\def\Bm      {{\ensuremath{\Bub}}\xspace}

\def\Bs      {{\ensuremath{\B^0_\squark}}\xspace}

\def\BsorBsbar {\kern \thebaroffset\optbar{\kern -\thebaroffset \Bs}\xspace}

\def\jpsi     {{\ensuremath{{\PJ\mskip -3mu/\mskip -2mu\Ppsi}}}\xspace}

\def\Y#1S{\ensuremath{\PUpsilon{(#1S)}}\xspace}

\def\LorLbar     {\kern \thebaroffset\optbar{\kern -\thebaroffset \PLambda}\xspace}

\def\BF         {{\ensuremath{\mathcal{B}}}\xspace}

\def\to                 {\ensuremath{\rightarrow}\xspace}

\def\CP                {{\ensuremath{C\!P}}\xspace}

\def\Vud  {{\ensuremath{V_{\uquark\dquark}^{\phantom{\ast}}}}\xspace}

\def\Vtd  {{\ensuremath{V_{\tquark\dquark}^{\phantom{\ast}}}}\xspace}

\def\Vubs  {{\ensuremath{V_{\uquark\bquark}^\ast}}\xspace}

\def\Vtbs  {{\ensuremath{V_{\tquark\bquark}^\ast}}\xspace}

\def\AT#1     {\ensuremath{A_{\mathrm{T}}^{#1}}\xspace}           % 2

\def\C#1      {\ensuremath{\mathcal{C}_{#1}}\xspace}                       % 9
\def\Cp#1     {\ensuremath{\mathcal{C}_{#1}^{'}}\xspace}                    % 7
\def\Ceff#1   {\ensuremath{\mathcal{C}_{#1}^{\mathrm{(eff)}}}\xspace}        % 9  
\def\Cpeff#1  {\ensuremath{\mathcal{C}_{#1}^{'\mathrm{(eff)}}}\xspace}       % 7
\def\Ope#1    {\ensuremath{\mathcal{O}_{#1}}\xspace}                       % 2
\def\Opep#1   {\ensuremath{\mathcal{O}_{#1}^{'}}\xspace}                    % 7

\newcommand{\aunit}[1]{\ensuremath{\text{\,#1}}}       
\newcommand{\tev}{\aunit{Te\kern -0.1em V}\xspace}
\newcommand{\gev}{\aunit{Ge\kern -0.1em V}\xspace}
\newcommand{\mev}{\aunit{Me\kern -0.1em V}\xspace}
\newcommand{\kev}{\aunit{ke\kern -0.1em V}\xspace}
\newcommand{\ev}{\aunit{e\kern -0.1em V}\xspace}
 
\newcommand{\mevc}{\ensuremath{\aunit{Me\kern -0.1em V\!/}c}\xspace}
\newcommand{\gevc}{\ensuremath{\aunit{Ge\kern -0.1em V\!/}c}\xspace}
\newcommand{\mevcc}{\ensuremath{\aunit{Me\kern -0.1em V\!/}c^2}\xspace}
\newcommand{\gevcc}{\ensuremath{\aunit{Ge\kern -0.1em V\!/}c^2}\xspace}
\def\fb   {\ensuremath{\aunit{fb}}\xspace}
\def\invfb   {\ensuremath{\fb^{-1}}\xspace}
\def\ab   {\ensuremath{\aunit{ab}}\xspace}

\def\gsim{{~\raise.15em\hbox{$>$}\kern-.85em
          \lower.35em\hbox{$\sim$}~}\xspace}
\def\lsim{{~\raise.15em\hbox{$<$}\kern-.85em
          \lower.35em\hbox{$\sim$}~}\xspace}

\def\tell1  {TELL1\xspace}
\def\ukl1   {UKL1\xspace}

\begin{document}
\title{\boldmath Rescaling the isospin triangle argument for constraining $\phi_2$ ($\alpha$)}
\subtitle{Consolidating Belle~II and a potential path forward for LHCb}
\author{J. Dalseno\inst{1}% etc
}                     % Do not remove
%
%%\offprints{}          % Insert a name or remove this line
%
\institute{Instituto Galego de F\'{i}sica de Altas Enerx\'{i}ıas (IGFAE), Universidade de Santiago de Compostela, Santiago de Compostela, Spain, \email{jeremy.peter.dalseno@cern.ch}}
\date{Received: 10 November 2021 / Accepted: 27 June 2022}
\abstract{
A rescaling of the SU(2) isospin triangles constraining $\phi_2$ ($\alpha$) that relies on measurements of the experimentally cleaner relative branching fractions, as opposed to those absolute, is proposed. Paving the way towards more systematically sustainable analysis, this method promises to eliminate a dominant systematic at Belle~II amongst others, namely the uncertainty on the number of $B \bar B$ pairs in data. Furthermore, a $\phi_2$ constraint in the $B \to \rho \rho$ system at LHCb that is more independent of Belle~II input is shown to become viable even without a measurement of $C\!P$ violation in $B^0 \to \rho^+\rho^-$.
\PACS{
      {PACS-key}{discribing text of that key}   \and
      {PACS-key}{discribing text of that key}
     } % end of PACS codes
} %end of abstract
\maketitle
\section{\boldmath Introduction}
\label{sec:intro}

Violation of the combined charge-parity symmetry~(\CP violation) in the Standard Model~(SM) arises from a single irreducible phase in the Cabibbo-Kobayashi-Maskawa~(CKM) quark-mixing matrix~\cite{Cabibbo:1963yz,Kobayashi:1973fv}. Various processes offer different yet complementary insight into this phase, which manifests in a number of experimental observables over-constraining the Unitarity Triangle (UT). The measurement of such parameters and their subsequent combination is important as New Physics~(NP) contributions can present themselves as an inconsistency within the triangle paradigm.
\begin{figure}[!htb]
    \centering
    \includegraphics[height=115pt,width=!]{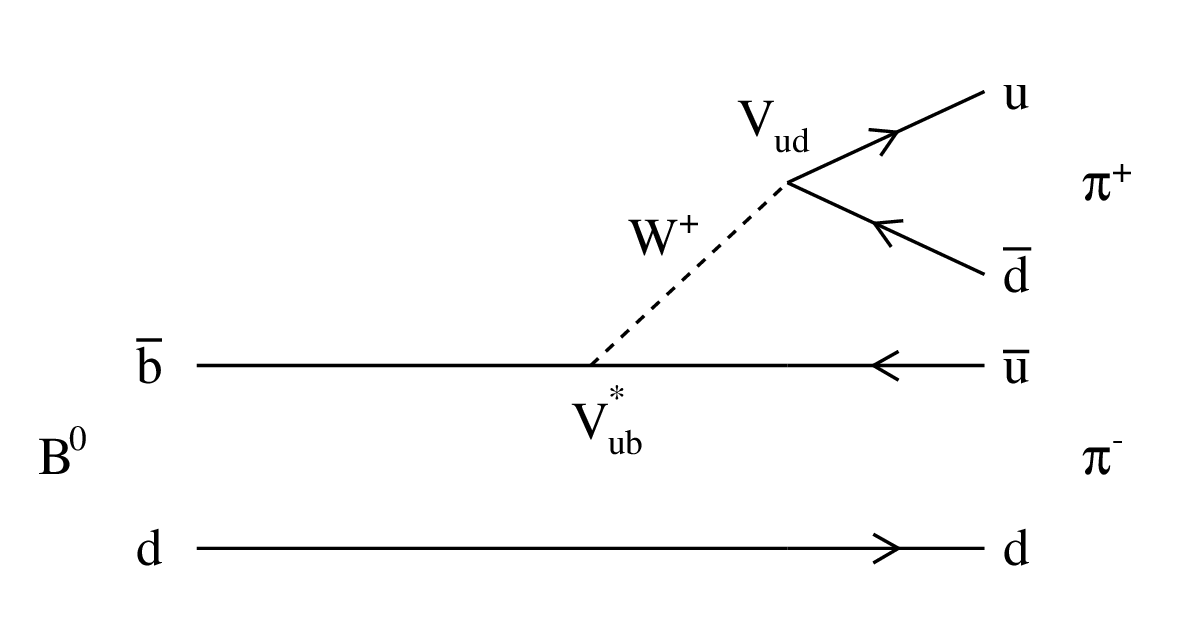}
    \put(-210,105){(a)}

    \includegraphics[height=115pt,width=!]{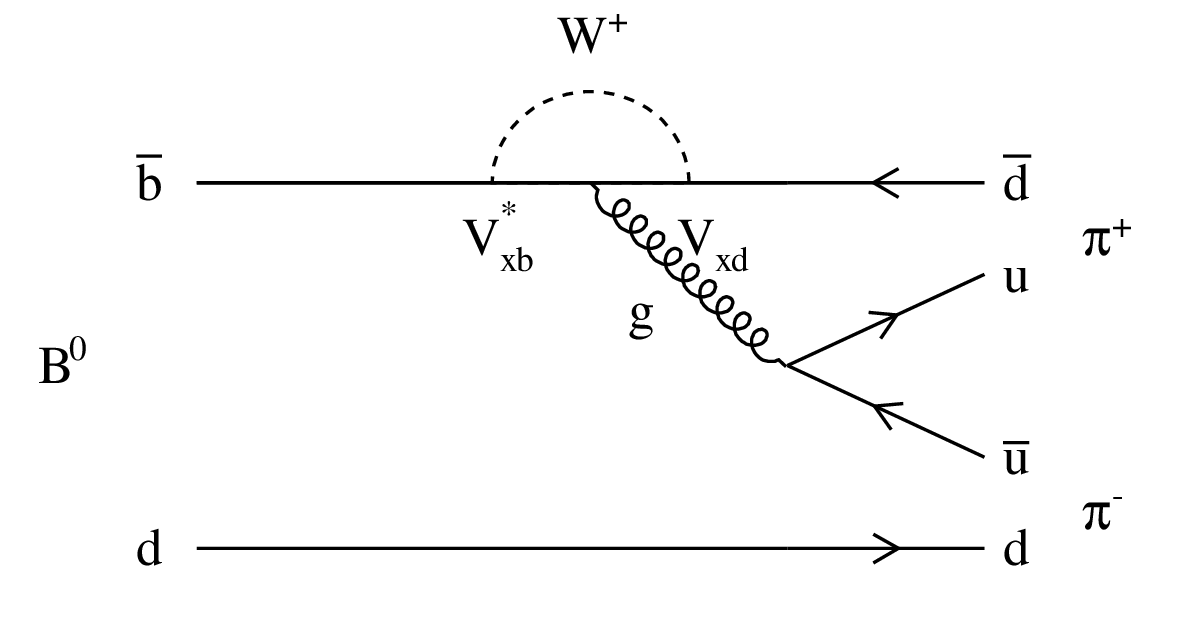}
%    \put(-425,105){(a)}
    \put(-210,105){(b)}

    \caption{\label{fig:feynman} Leading-order Feynman diagrams shown producing $\Bz \to \pip \pim$ decays, though the same quark transition can also produce $\Bz \to \rho^{\pm} \pimp$, $\rho^+ \rho^-$ and $a_1^\pm \pimp$. (a) depicts the dominant (tree) diagram while (b) shows the competing loop (penguin) diagram. In the penguin diagram, the subscript $x$ in $V_{xb}$ refers to the flavour of the intermediate-state quark $(x=u,c,t)$.}
\end{figure}

Decays that proceed predominantly through the $\bquarkbar \to \uquarkbar \uquark \dquarkbar$ tree transition (Fig.~\ref{fig:feynman}a) in the presence of \Bz--\Bzb mixing are sensitive to the interior angle of the UT, $\phi_2 = \alpha \equiv \arg(-\Vtd \Vtbs)/(\Vud \Vubs)$, which can be accessed through mixing-induced \CP violation observables measured from time-dependent, flavour-tagged analyses. This quark
process manifests itself in multiple systems, including $B \to \pi \pi$~\cite{Lees:2012mma,Adachi:2013mae,Aaij:2018tfw,Aaij:2020buf,Aubert:2007hh,Duh:2012ie,Julius:2017jso}, $(\rho \pi)^0$~\cite{Lees:2013nwa,Kusaka:2007dv,Kusaka:2007mj}, $\rho \rho$~\cite{Aubert:2007nua,Vanhoefer:2015ijw,Aubert:2009it,Zhang:2003up,Aubert:2008au,Adachi:2012cz,Aaij:2015ria} and $a_1^\pm \pi^\mp$~\cite{Aubert:2006gb,Dalseno:2012hp,Aubert:2009ab}, where the angle $\phi_2$ has so far been constrained with an overall uncertainty of around $4^\circ$~\cite{Gronau:2016idx,Charles:2017evz,Bona:2006ah,Amhis:2019ckw}. With the dubious honour of being the least known input to the UT now falling to $\phi_2$, there has never been better motivation to improve its experimental precision.

One significant limitation hindering this endeavour is the presence of dominant irreducible systematic uncertainties plaguing absolute branching fraction measurements of the decay processes involved during the procedure to remove bias induced by amplitudes competing with the primary tree diagram. In consequence, current procedure would negate the vast improvement in statistical precision expected over the coming years at \belletwo as an undesirable side effect. Furthermore, \lhcb inherits the total uncertainties on normalising branching fractions from \belletwo as their baseline, rendering their $\phi_2$ programme noncompetitive from the onset. Through a reexamination of the isospin triangle relations that constrain $\phi_2$, this paper demonstrates how to eliminate this issue from consideration.

I open in Sect.~\ref{sec:isospin}, with a description of the SU(2)-based approach for controlling distortions in experimental $\phi_2$ measurements arising from the ever-present strong-loop gluonic penguin processes. Following this, I introduce in Sect.~\ref{sec:rescaled} a rescaling of the isospin triangle argument, improving overall experimental precision in $\phi_2$ and relieving \lhcb of their dependency on \belletwo. The impact of this modification at \belletwo is studied for the $B \to \pi \pi$ system, then for the $B \to \rho \rho$ system at \lhcb before finally, conclusions are drawn in Sect.~\ref{sec:conclusion}.

%\section{\boldmath Strong-penguin containment in \texorpdfstring{$\phi_2$}{phi2} constraints}
\section{\boldmath Strong-penguin containment in $\phi_2$ constraints}
\label{sec:isospin}

In general, the extraction of $\phi_2$ is complicated by the presence of interfering amplitudes that distort the experimentally determined value of $\phi_2$ from its SM expectation and would mask any NP phase if not accounted for. These effects primarily include $\bquarkbar \rightarrow \dquarkbar u \uquarkbar$ strong-loop decays (Fig.~\ref{fig:feynman}b), although isospin-violating processes such as electroweak penguins, $\piz$--$\eta$--$\eta^\prime$ mixing, $\rho^0$--$\omega$--$\phi$ mixing~\cite{Gronau:2005pq}, the finite $\rho$ width in $B \to \rho\rho$~\cite{Falk:2003uq} and neglected systematic correlations arising from the $\rho$ pole parameters~\cite{Dalseno:2021bin} can also play a role.

\subsection{Original SU(2) approach}

It is possible to remove the isospin-conserving component of this contamination by invoking SU(2) arguments. The original method considers the three possible charge configurations of $B \rightarrow \pi\pi$ decays~\cite{Gronau:1990ka}. Bose-Einstein statistics rules out a total isospin $I=1$ contribution, leaving just the $I=0, 2$ amplitudes remaining. Strong penguins then only have the possibility to contribute an $I=0$ amplitude, since the mediating gluon is an isospin singlet. However, in the specific case of $\Bp \rightarrow \pip \piz$, the further limiting projection $I_{3} = 1$ additionally rules out $I=0$, thereby forbidding strong penguin contributions to this channel.
\begin{figure}[!htb]
    \centering
    \includegraphics[height=120pt,width=!]{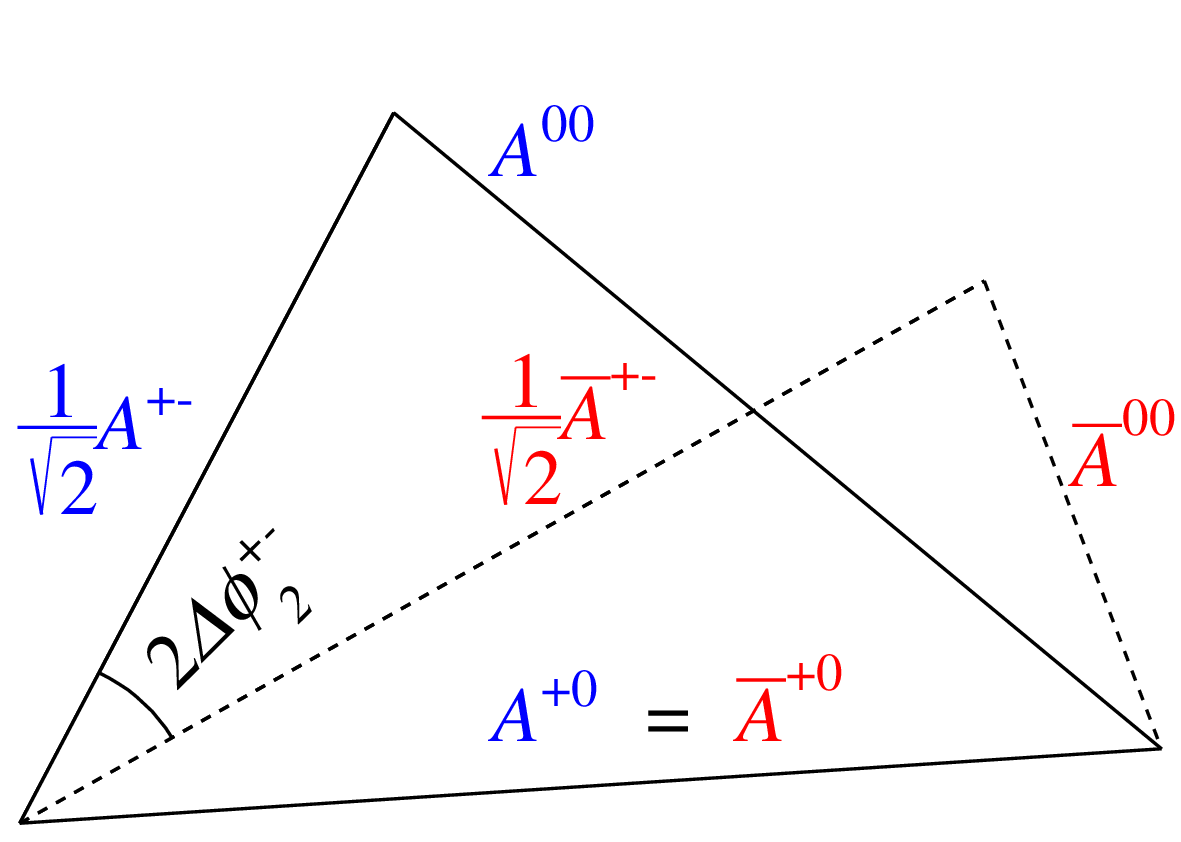}
    \caption{\label{fig_iso} Complex isospin amplitude triangles from which $\Delta \phi^\pm_2$ can be determined.}
\end{figure}

The complex $B \rightarrow \pi\pi$ and $\bar B \rightarrow \pi\pi$ decay amplitudes obey the isospin relations
\begin{equation}
    A^{+0} = \frac{1}{\sqrt{2}}A^{+-} + A^{00}, \;\;\;\; \bar{A}^{+0} = \frac{1}{\sqrt{2}}\bar{A}^{+-} + \bar{A}^{00},
  \label{eq_iso}
\end{equation}
respectively, where the superscripts refer to the combination of pion charges. The decay amplitudes can be represented as triangles in the complex plane as shown in Fig.~\ref{fig_iso}. As $\Bp \rightarrow \pip \piz$ is a pure tree mode, its amplitude in isospin space is identical to that of its \CP-conjugate and so these triangles lose their relative orientation to share the same base, $A^{+0}=\bar{A}^{+0}$, allowing the shift in $\phi_2$ caused by strong penguin contributions $\Delta \phi^\pm_2 \equiv \phi_2^\pm - \phi_2$, to be determined from the phase difference between $\bar{A}^{+-}$ and $A^{+-}$. These amplitudes can be constrained by seven mostly independent physical observables for a two-fold discrete ambiguity in the range $[0, 180]^\circ$, which are related to the decay amplitudes as
%\begin{equation}
%    \label{eq:old}
%    \frac{1}{\tau_B^{i+j}} \BF^{ij} = \frac{|\bar A^{ij}|^2 + |A^{ij}|^2}{2}, \hspace{10pt} \mathcal{A}_{\CP}^{ij} = \frac{|\bar A^{ij}|^2 - |A^{ij}|^2}{|\bar A^{ij}|^2 + |A^{ij}|^2}, \hspace{10pt} \mathcal{S}_{\CP}^{ij} = \frac{2\Im(\bar A^{ij} A^{ij*})}{|\bar A^{ij}|^2 + |A^{ij}|^2},
%\end{equation}
\begin{eqnarray}
    \label{eq:old}
    \frac{1}{\tau_B^{i+j}} \BF^{ij} &=& \frac{|\bar A^{ij}|^2 + |A^{ij}|^2}{2}, \nonumber\\
    \mathcal{A}_{\CP}^{ij} &=& \frac{|\bar A^{ij}|^2 - |A^{ij}|^2}{|\bar A^{ij}|^2 + |A^{ij}|^2}, \nonumber\\
    \mathcal{S}_{\CP}^{ij} &=& \frac{2\Im(\bar A^{ij} A^{ij*})}{|\bar A^{ij}|^2 + |A^{ij}|^2},
\end{eqnarray}
where \BF, $\mathcal{A}_{\CP}$ and $\mathcal{S}_{\CP}$ are the branching fractions, \CP violation in the decay and mixing-induced \CP violation parameters, respectively. The mixing-induced \CP-violating parameter implicitly assumes the mixing phase is absorbed into the decay amplitude and that there is no \CP violation in mixing, $|q/p|=1$. The superscript $ij$, represents the charge configuration of the final state pions and $\tau_B^{i+j}$ is the lifetime of the \Bp~($i+j=1$) or \Bz~($i+j=0$). Naturally for $\Bp \rightarrow \pip \piz$, \CP violation in the decay is forbidden by the isospin argument and mixing-induced \CP violation is not defined. The ambiguity in $\phi_2$ is also increased to eight-fold if $\mathcal{S}_{\CP}^{00}$ of the colour-suppressed channel is not measured as is currently the case for $B \to \pi\pi$ decays. This approach can also be applied to the $B \to \rho \rho$ system analogously, substituting the $\rho$ meson in place of each pion.

\subsection{Next-generation approach}

However, the $B \to \rho \rho$ system presents a greater theoretical and experimental challenge over $B \to \pi \pi$. It has already been pointed out that isospin-breaking ($I=1$) $\rho$-width effects can be controlled by reducing the $\rho$ analysis window of $\Bz \to \rho^+ \rho^-$ and $\Bp \to \rho^+ \rho^0$ according to the method outlined in Ref.~\cite{Gronau:2016nnc}. An open question to be studied is the extent to which this is systematically feasible in the presence of interfering and non-interfering backgrounds.

I espouse an alternate viewpoint in which the possibility to exploit the multi-body final state through directly modelling the structure of $\rho^0$--$\omega$ mixing and $I=1$ finite $\rho$-width effects is acquired in exchange for greater analysis complexity. To that end, I have already outlined the amplitude analysis framework by which this can be achieved, replacing the measured \CP-violation observables from Eq.~\ref{eq:old} by
\begin{equation}
    \label{eq:new}
    |\lambda_{\CP}^{ij}| = \biggl|\frac{\bar A^{ij}}{A^{ij}}\biggr|, \hspace{10pt} \phi_2^{ij} = \frac{\arg(\bar A^{ij} A^{ij*})}{2},
\end{equation}
where $\lambda^{ij}_{\CP}$ is a \CP-violation parameter and $\phi_2^{ij}$ is its effective weak phase. As these quantities are now related to the isospin triangles at amplitude level, the solution degeneracy in $\phi_2$ for the range $[0, 180]^\circ$ is resolved~\cite{Dalseno:2018hvf} and as an added incentive, the eight-fold solution degeneracy in $B^0 \to a_1^\pm \pimp$ can also be lifted for the same range in the SU(3) approach with the added possibility to directly constrain non-factorisable symmetry-breaking effects~\cite{Dalseno:2019kps}.

\subsection{Statistical method}

In this paper, I employ the frequentist approach adopted by the CKMfitter Group~\cite{Charles:2017evz} where a $\chi^2$ is constructed comparing theoretical forms for physical observables expressed in terms of parameters of interest, $\bm{\mu}$, with their experimentally measured values, $\bm{ x}$. The most general form,
\begin{equation}
    \chi^2 \equiv (\bm{x}-\bm{\mu})^T \bm{\Sigma}^{-1} (\bm{x}-\bm{\mu}),
\end{equation}
is necessary here, where $\bm{\Sigma}$ is total covariance matrix composed of the statistical and systematic covariance matrices as $\bm{\Sigma} \equiv \bm{\Sigma}_{\rm Stat} + \bm{\Sigma}_{\rm Syst}$. The statistical covariance matrix comes directly from the function minimisation procedure during the nominal fit to a sample, while the systematic covariance matrix is manually derived. Parameter variations are generated according to their uncertainties and the fit is repeated for each set of variations. Over $N$ fits, the covariance between a pair of physical observables is given by
\begin{equation}
    \Sigma_{x,y} \equiv \sum^N_{i=1} \frac{(x_i-\bar x)(y_i-\bar y)}{N},
\end{equation}
where the barred quantities representing the means are obtained from the nominal fit.

A scan is then performed, minimising the $\chi^2$ to determine $\bm{\mu}$ for each value of $\phi_2$ fixed across a range. The value of $\Delta \chi^2$ from the global minimum is finally converted into a $p$-value scan, assuming it is distributed with one degree of freedom, from which confidence intervals can be derived.

\section{Rescaled isospin triangles}
\label{sec:rescaled}

\begin{figure}
    \centering
    \includegraphics[height=120pt]{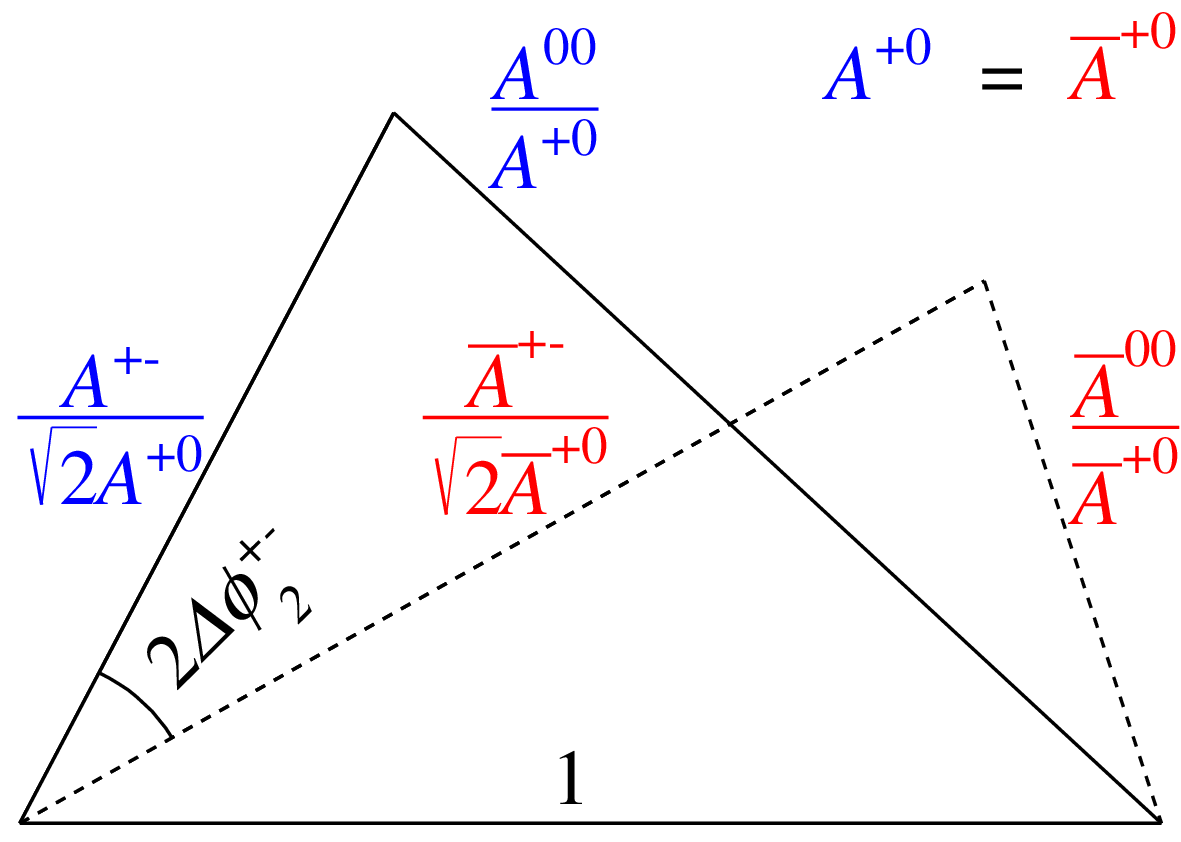}
    \caption{\label{fig_iso_rs} Rescaled complex isospin amplitude triangles from which $\Delta \phi^\pm_2$ can be determined.}
\end{figure}

In the current paradigm, absolute branching fractions provide knowledge on the base length of the isospin triangles to constrain $\Delta \phi_2^\pm$, which could be considered a nuisance parameter for the purpose of measuring $\phi_2$ and in subsequent fits of the CKM matrix. As only two independent parameters are required to constrain a triangle, a rescaling to share a base length of unity is envisioned as shown in Fig.~\ref{fig_iso_rs}, reminiscent of the treatment of the UT. This reduces the number of required observables also by one and the experimentally cleaner ratio of branching fractions may instead be used in the constraint. While theoretical uncertainties arising from isospin-breaking effects should not be adversely impacted through the constraint of amplitude ratios, the question if any significant improvements are possible through this approach is left to the theoretical community.

Once again following the parametrisation from the CKMfitter Group, the isospin triangle amplitudes may be conveniently composed as
\begin{align}
    \label{eq:rs}
    A^{+0} &= \mu e^{i(\Delta - \phi_2)},& \bar A^{+0} &= \mu e^{i(\Delta + \phi_2)}, \nonumber \\ 
    A^{+-} &= \mu a,& \bar A^{+-} &= \mu \bar a e^{2i\phi_2^\pm},
\end{align}
where $\mu$ is now fixed to unity and only $a$, $\bar a$, $\Delta$ and $\phi_2^\pm$ are free parameters of the model. To constrain these parameters, the two necessary experimental observables related to the branching fractions become $\BF^{+-}/\BF^{+0}$ and $\BF^{00}/\BF^{+0}$, constructed as
\begin{equation}
    \frac{\tau_\Bp \BF^{ij}}{\tau_\Bz \BF^{+0}}  = \frac{|\bar A^{ij}|^2 + |A^{ij}|^2}{|\bar A^{+0}|^2 + |A^{+0}|^2},
\end{equation}
while the measured \CP-violating parameters remain as per their usual construction given in Eqs.~\ref{eq:old} or~\ref{eq:new}.

%\subsection{\texorpdfstring{$B \to \pi \pi$}{B to pi pi} at \belletwo}
\subsection{$B \to \pi \pi$ at \belletwo}
\label{sec:pipi}

An immediate benefit at Belle II would be the elimination of an otherwise irreducible systematic uncertainty related to the number of $B \Bbar$ pairs produced at $\epem$ machines operating at the $\Upsilon(4S)$ resonance, $N_{B\Bbar}$, which was found to exceed 1\% at both \babar and \belle. This quantity enters the absolute branching fraction calculations in $B \to \pi \pi$ decays through
\begin{equation}
    \BF^{ij} = \frac{N^{ij}}{\epsilon^{ij}N_{B\Bbar}},
\end{equation}
where $N^{ij}$ is the extracted signal yield and $\epsilon^{ij}$ is the reconstruction efficiency of that mode. As an aside, although equal production of \Bp \Bm and \Bz \Bzb pairs is implicitly assumed here for simplicity, this will need to be evaluated at \belletwo  without bias~\cite{Jung:2015yma} as the current uncertainties on their rates~\cite{ParticleDataGroup:2020ssz} would otherwise constitute the dominant systematic instead of those arising from $N_{B\Bbar}$.

To illustrate the potential improvement through the measurement of relative branching fractions, I repeat the $\phi_2$ projection for $50 \ab^{-1}$ at \belletwo, comparing with the current approach based on absolute branching fractions and the enhanced approach of Ref.~\cite{Dalseno:2021bin}, where systematic correlations in $N_{B\Bbar}$ are taken into account. Input is borrowed from Ref.~\cite{Kou:2018nap} and manipulated to instead provide a projection on the ratio of branching fractions as given in Table~\ref{tab:pipi}. To facilitate a fair comparison between all 3 approaches, it will be assumed that the systematic uncertainty is wholly due to the uncertainty in $N_{B\Bbar}$. This means that in the rescaled approach, the systematic uncertainty now derives from the statistical uncertainty of $\BF^{+0}$, also requiring both ratios of branching fractions to be fully systematically correlated.
%Their total systematics are calculated by quadratically subtracting the $N_{B\Bbar}$ uncertainty of $1.37\%$ found at \belle from the absolute branching fraction errors and propagating the remainders in the usual way. For the correlated systematics approach with absolute branching fractions, the systematics covariance matrix can also be calculated in a similar way through subtraction of the relevant covariance matrices, with the correlation matrix for $N_{B\Bbar}$ taken from Ref.~\cite{Dalseno:2021bin}.

\begin{table}[!htb]
    \centering
    \caption{\label{tab:pipi} Projections for $B \to \pi \pi$ physics observables with $50 \ab^{-1}$ adapted from Ref.~\cite{Kou:2018nap} where the first uncertainty is statistical and the second is systematic.}
%    \begin{tabular}{|c|c|}
    \begin{tabular*}{\columnwidth}{@{\extracolsep{\fill}}lc@{}}
    \hline
    Parameter & \belletwo projection\\ \hline
%    $\BF^{+-}/\BF^{+0}$ & $\phantom{+}0.860 \pm 0.007 \pm 0.009$\\
%    $\BF^{00}/\BF^{+0}$ & $\phantom{+}0.224 \pm 0.005 \pm 0.004$\\
    $\BF^{+-}/\BF^{+0}$ & $\phantom{+}0.860 \pm 0.005 \pm 0.004$\\
    $\BF^{00}/\BF^{+0}$ & $\phantom{+}0.224 \pm 0.005 \pm 0.001$\\
    $\mathcal{A}^{+-}_{\CP}$ & $+0.33\phantom{0} \pm 0.01\phantom{0} \pm 0.03\phantom{0}$\\
    $\mathcal{S}^{+-}_{\CP}$ & $-0.64\phantom{0} \pm 0.01\phantom{0} \pm 0.01\phantom{0}$\\
    $\mathcal{A}^{00}_{\CP}$ & $+0.14\phantom{0} \pm 0.03\phantom{0} \pm 0.01\phantom{0}$\\ \hline
    %\end{tabular}
%    \caption{\label{tab:pipi} Projections for $B \to \pi \pi$ physics observables with $50 \ab^{-1}$ adapted from ref.~\cite{Kou:2018nap} where the first uncertainty is statistical and the second is systematic.}
    \end{tabular*}
\end{table}

\begin{figure}[!htb]
    \centering
    \includegraphics[height=135pt,width=!]{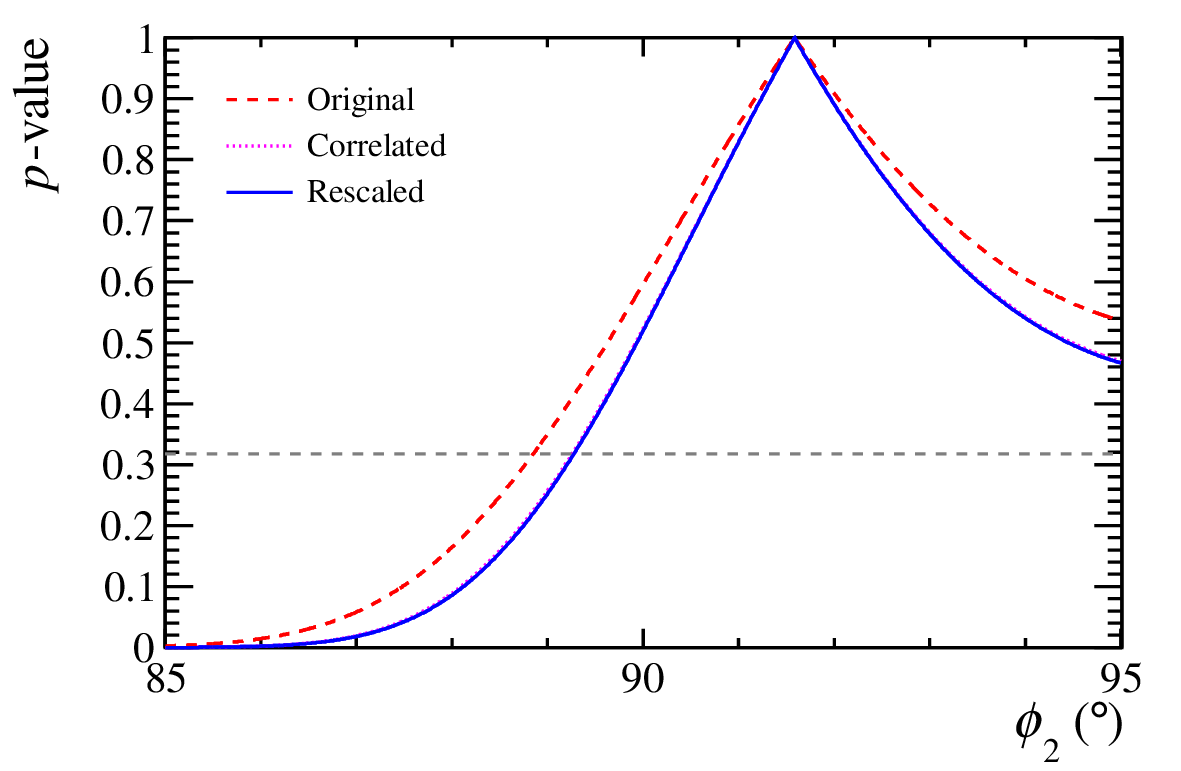}
    \put(-30,115){(a)}

    \includegraphics[height=135pt,width=!]{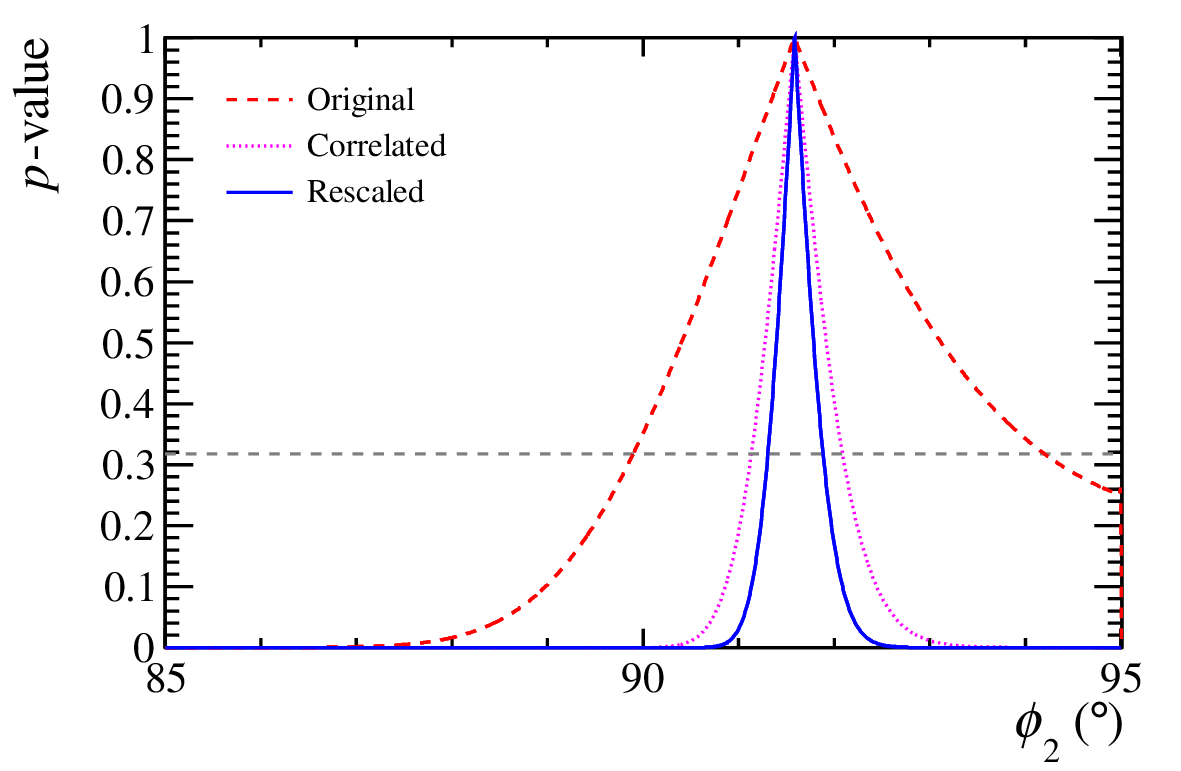}
    \put(-30,115){(b)}

    \caption{\label{fig:pipi} $p$-value scans of $\phi_2$ under (a) \belletwo conditions and (b)~with 10 times more data, where the horizontal dashed line indicates the $1\sigma$ bound. The dashed red curve shows the original approach conducted with absolute branching fractions, the dotted magenta considers systematic correlations in $N_{B\Bbar}$, while the solid blue relies on measurements of the ratios of branching fractions.}
\end{figure}

As can be seen from the results shown in Fig.~\ref{fig:pipi}a, the leading edge of the solution consistent with the SM is seen to improve by $0.4^\circ$ for the correlated systematics approach with absolute branching fractions and a further $0.02^\circ$ when working with the relative branching fractions. These are striking improvements within the context of the sub-degree precision anticipated at \belletwo, although this should be considered an idealised scenario. In reality, the cancellation or reduction in other leading detection efficiency systematics from the common pion in each ratio is conceivable, while others may be completely uncorrelated. While improvements with the rescaled approach should be marginal over accounting for just the $N_{B\Bbar}$ systematic correlations with absolute branching fractions at \belletwo, these differences can only become more pronounced when considered in light of a potential upgrade. Assuming an order of magnitude more data in a hypothetical \belletwo upgrade scenario, all variances are scaled accordingly save for those relating to the irreducible $N_{B\Bbar}$ systematic, the results of which can be seen in Fig.~\ref{fig:pipi}b. As considering systematic correlations in the absolute branching fractions approach can never completely remove their impact in the presence of other uncertainties, the rescaled proposition should ultimately be preferred as it provides superior performance in the long term, creating experimental conditions more conducive towards systematically sustainable analysis.

%Original
%1-CL = 0.317311: 1.161, 88.839, 102.087, 120.249, 133.533, 136.467, 149.751, 167.913
%Systematic correlations
%1-CL = 0.317311: 0.891, 89.109, 101.961, 120.357, 133.227, 136.773, 149.643, 168.039
%Rescaled
%1-CL = 0.317311: 0.855, 89.145, 101.943, 120.375, 133.173, 136.827, 149.625, 168.057
%
%Pure NBB systematic
%1-CL = 0.317311: 89.2545
%1-CL = 0.317311: 89.2735
%
%       1 0.516335 0.769904        0        0        0
%0.516335        1 0.533637        0        0        0
%0.769904 0.533637        1        0        0        0
%       0        0        0        1        0        0
%       0        0        0        0        1        0
%       0        0        0        0        0        1
%
%Belle III, scale by factor of 10
%1-CL = 0.317311: 89.8945, 94.2015
%1-CL = 0.317311: 91.1315, 92.0855
%1-CL = 0.317311: 91.3065, 91.8805
%\subsection{\texorpdfstring{$B \to \rho \rho$}{B to rho rho} at \lhcb}
\subsection{$B \to \rho \rho$ at \lhcb}
\label{sec:rhorho}

While this improvement naturally applies to the $B \to \rho \rho$ system at \belletwo in addition, a more intriguing implication arises at \lhcb, opening an opportunity to extend their physics reach outlined in Ref.~\cite{LHCb:2018roe}. Clearly, branching fractions measured relative to $\Bp \to \rhop \rhoz$ negates the need to borrow a normalisation channel from \belletwo, removing the stifling propagation of their total uncertainties that this entails. Moreover, there are additional competing pressures. While \lhcb has long been able to handle \piz reconstruction albeit with more difficulty, the long-standing consensus is that even if reconstruction with two neutral pions were to succeed, a subsequent need for the relatively inefficient flavour tagging in an hadronic initial-state environment to obtain the \CP-violation parameters of $\Bz \to \rhop \rhom$ lands the knockout blow. 

However, it is important to remember the history of the $B \to \rho \rho$ constraint on $\phi_2$. Recall that despite no measurement of \CP violation in the decay of $\Bz \to \rhoz \rhoz$ from Belle for which flavour tagging would have been required, a good constraint on $\phi_2$ is nevertheless achieved~\cite{Vanhoefer:2015ijw}. This is due to its small branching fraction relative to the colour-favoured processes which renders the isospin triangles essentially flat, limiting the impact of a \CP-violation measurement for this channel in the first instance. Therefore, given isospin triangles of the same orientation, the converse of this scenario must also be true. If the \CP-violation parameters of $\Bz \to \rhoz \rhoz$ could be precisely determined as would be the case at \lhcb, then the \CP-violation parameters of $\Bz \to \rhop \rhom$ are likewise unnecessary, removing the penalty from their flavour-tagging performance. In such a scenario, sensitivity to $\phi_2$ derives purely from the measurement of $\mathcal{S}_{\CP}^{00}$ as opposed to $\mathcal{S}_{\CP}^{+-}$ in the original approach.

The limiting factor at \lhcb will obviously come from the yield of $\Bz \to \rhop \rhom$. As an estimate of the $\Bp \to \rhop \rhoz$ yield is formally unavailable at this time, a scan of the potential total error space in $\BF^{+-}/\BF^{+0}$ and $\BF^{00}/\BF^{+0}$ is performed to understand the feasibility of this approach. Input for this study is based on the final \babar results~\cite{Aubert:2007nua,Aubert:2009it,Aubert:2008au} and for simplicity only assumes the longitudinal polarisation as recorded in Table~\ref{tab:rhorho}. Uncertainties are considered with 23\invfb and 300\invfb to be collected by \lhcb after the Run~3 and Run~5 data taking periods, respectively. The total uncertainties on the ratios of branching fractions correspond to an effective yield range in the least known channel from as little as 100 events to a more ambitious 10~000 in Run~3, while for Run~5, the relative branching fractions are assumed to be systematically limited at best, for a total uncertainty of 0.5\% taken from a high-statistics \lhcb analysis performed with 3-body charmless hadronic final states~\cite{LHCb:2020xcz}. Given that roughly half a million four-charged-pion events are expected to pass selection in Run~3 alone as estimated in a previous study~\cite{Dalseno:2018hvf}, itself based on Ref.~\cite{Aaij:2015ria}, the worst-case effective yield for $\Bz \to \rhop \rhom$, which has a similar branching fraction, seems very reasonable. The \CP-violation parameter uncertainties of $\Bz \to \rhoz \rhoz$ are taken from that same study and scaled to Run~5 as needed. With these levels of effective yields, a rudimentary time-dependent analysis of $\Bz \to \rhop \rhom$ with large uncertainties may become possible by the end of Run~5 and is included as a separate consideration. These \CP-violation parameter uncertainties scale with the $\BF^{+-}/\BF^{+0}$ error based on a Run~1 \lhcb study of $\Bz \to \jpsi \KS$ with an effective yield which happens to have a similar uncertainty of order 1\%~\cite{LHCb:2017mpa}.

%\begin{table}[!htb]
%    \centering
%    \begin{tabular}{|c|c|c|c|} \hline
\begin{table*}[!htb]
    \centering
    \caption{\label{tab:rhorho} Total uncertainty projections for $B \to \rho \rho$ physics observables with Run~3 and Run~5.}
    \begin{tabular*}{\textwidth}{@{\extracolsep{\fill}}lccc@{}}
    \hline
    Parameter & Central value~\cite{Aubert:2007nua,Aubert:2009it,Aubert:2008au} & Run 3 uncertainty & Run 5 uncertainty\\ \hline
    $\BF^{+-}/\BF^{+0}$ & $\phantom{+}1.12$ & [1, 10]\% & [0.5, 2]\%\\
    $\BF^{00}/\BF^{+0}$ & $\phantom{+}0.03$ & [1, 10]\% & [0.5, 2]\%\\
    $|\lambda^{00}_{\CP}|$ & $\phantom{+}0.82$ & $0.04$~\cite{Dalseno:2018hvf} & $0.01$\\
    $\phi_2^{00} \; ({}^\circ)$ & $\phantom{+}80.9$ & $2.0\phantom{0}$~\cite{Dalseno:2018hvf} & $0.5\phantom{0}$\\ \hline
    $\mathcal{A}^{+-}_{\CP}$ & $-0.01$ & --- & [0.04, 0.14]~\cite{LHCb:2017mpa}\\
    $\mathcal{S}^{+-}_{\CP}$ & $-0.17$ & --- & [0.04, 0.16]~\cite{LHCb:2017mpa}\\ \hline
%    \end{tabular}
%    \caption{\label{tab:rhorho} Total uncertainty projections for $B \to \rho \rho$ physics observables with Run~3 and Run~5.}
%\end{table}
    \end{tabular*}
\end{table*}

\begin{figure}[!htb]
    \centering
    \includegraphics[height=135pt]{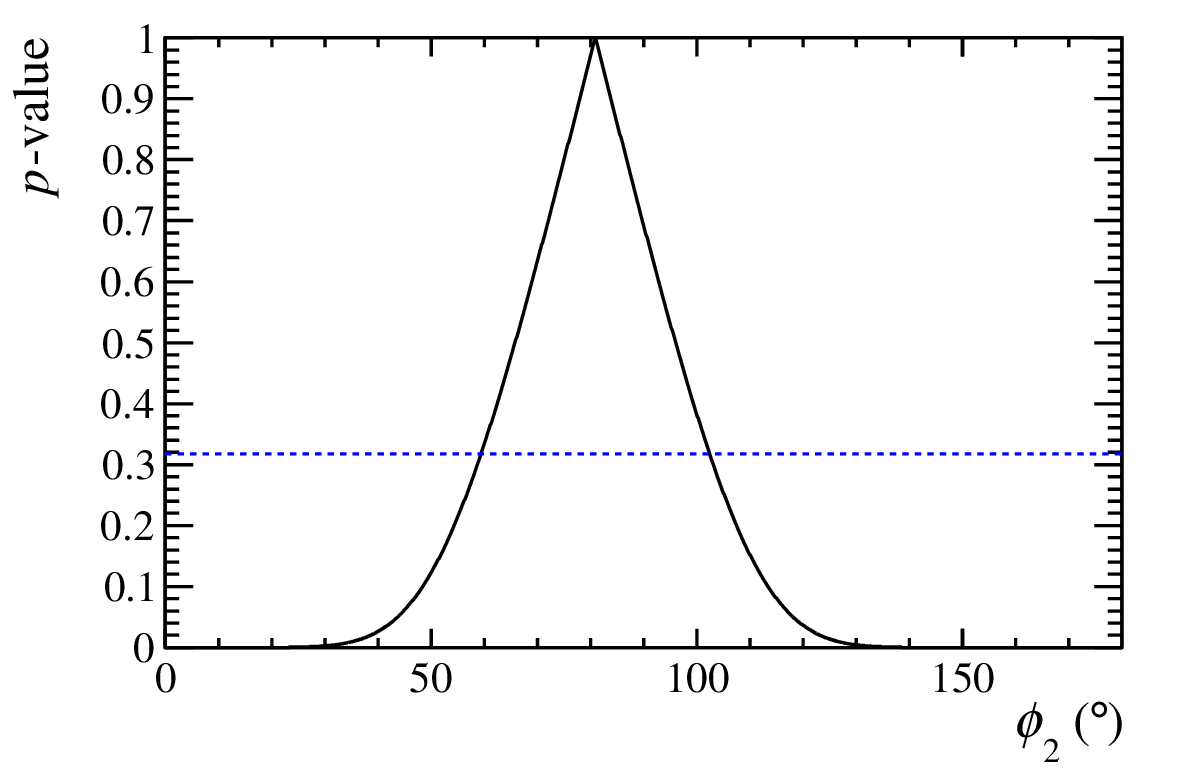}
    \put(-170,115){(a)}

    \includegraphics[height=135pt]{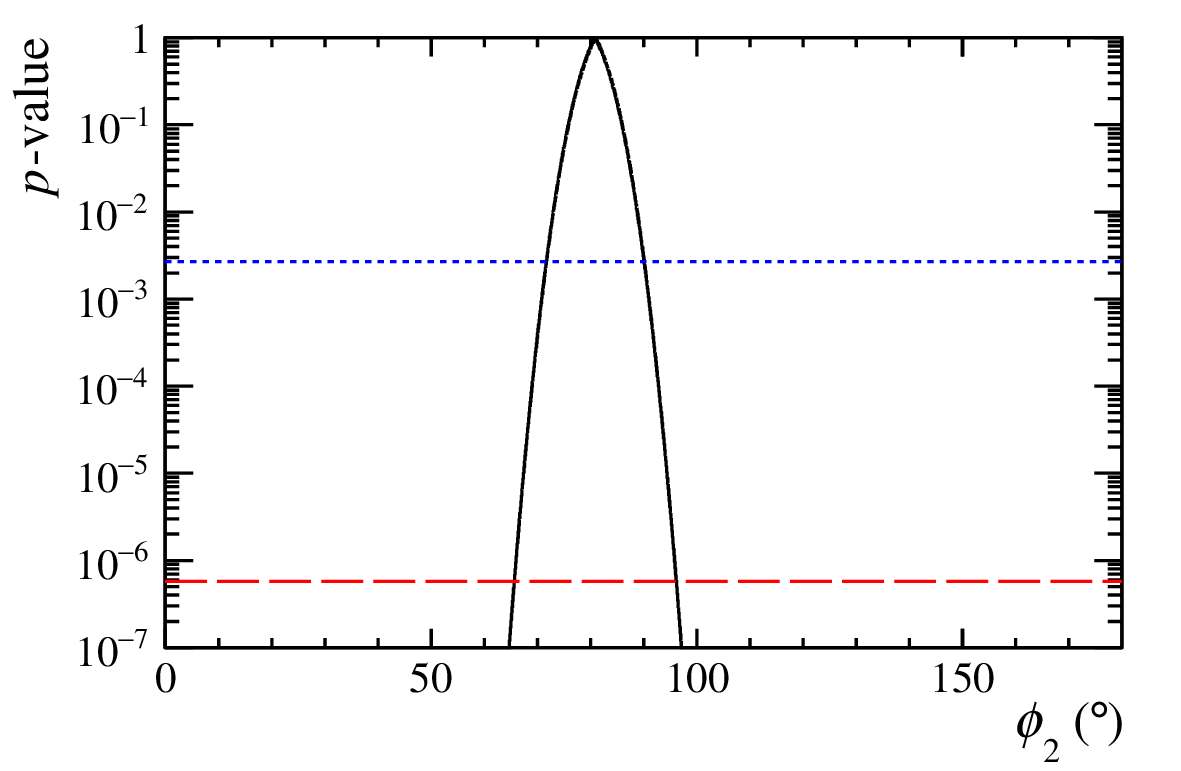}
%    \put(-380,115){(a)}
    \put(-170,115){(b)}

    \caption{\label{fig:rhorhoscenario} $p$-value scans of $\phi_2$ for the (a)~worst and (b)~best-case scenarios in Run~3. In the worst case, the horizontal dotted blue line shows the $1\sigma$ bound, while for the best case, it is more instructive to show the $3\sigma$ and $5\sigma$ bounds as the blue-dotted and red-dashed lines, respectively.}
\end{figure}

A sanity check showing the $\phi_2$ constraint for the Run~3 best and worst-case scenarios corresponding to total uncertainties in the relative branching fractions of 1\% and 10\%, respectively, can be seen in Fig.~\ref{fig:rhorhoscenario}. With now only four physics observables to constrain the isospin triangles and $\phi_2$, there are indeed no surprises found lurking in the range, $[0,180]^\circ$. In principle, there should be a plateau at the apex of these scans similar to that seen in the very first constraint at \belle~\cite{Belle:2008slh}, as it is impossible to resolve the space between the four-fold ambiguity in triangle orientation that is present with this approach at \lhcb.

\begin{figure}[!htb]
    \centering
    \includegraphics[height=135pt]{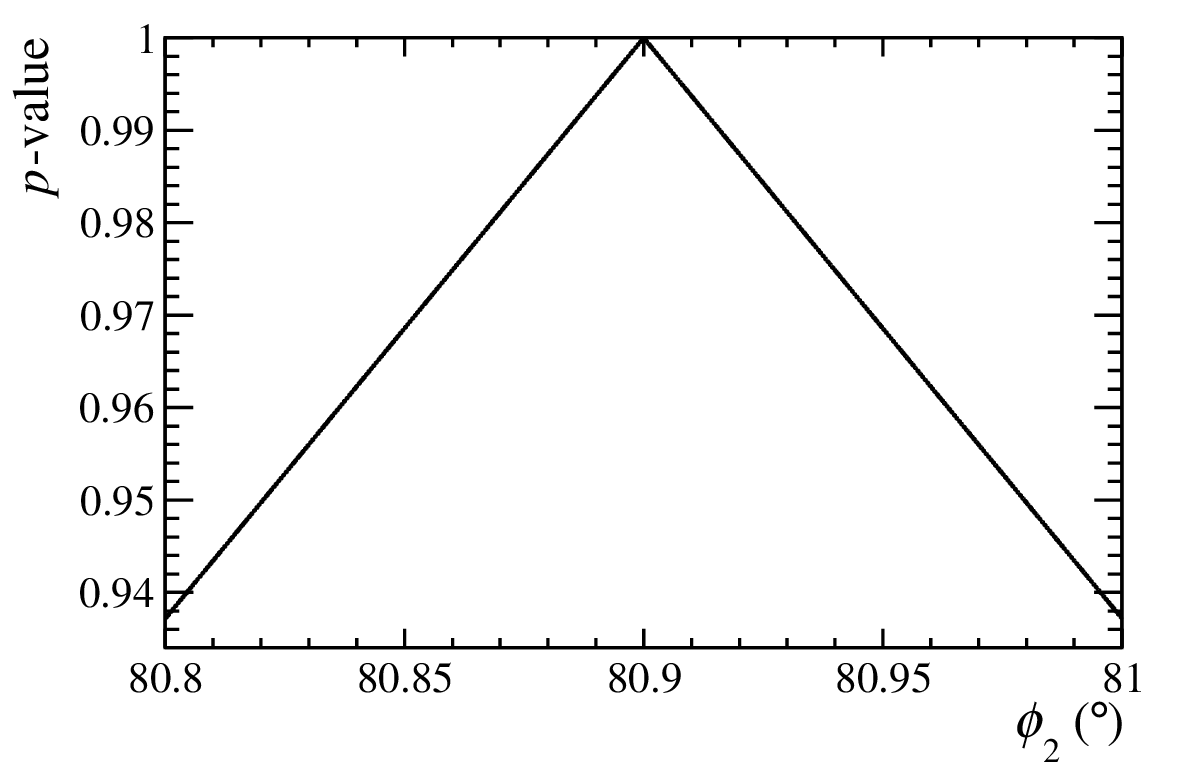}
    \put(-170,115){(a)}

    \includegraphics[height=135pt]{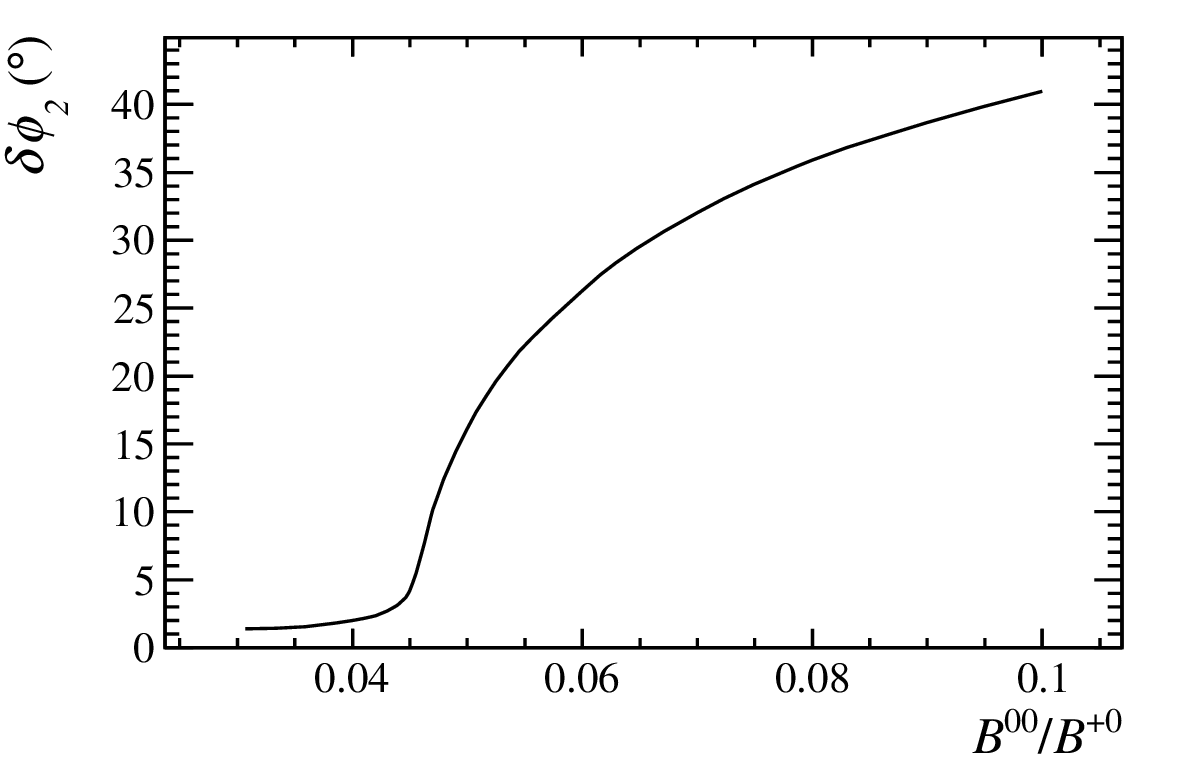}
%    \put(-380,115){(a)}
    \put(-170,115){(b)}

    \caption{(a) Zoomed in $p$-value scan of $\phi_2$ for the Run~5 best-case scenario without \CP-violation input from $\Bz \to \rhop \rhom$, while (b) tracks the $\phi_2$ uncertainty as a function of $\BF^{00}/\BF^{+0}$.}
    \label{fig:rhorhocheck}
\end{figure}

Zooming in towards the apex with a high fidelity scan of $10^{-5}$ degrees for the more extreme Run~5 conditions with no \CP-violation measurement in $\Bz \to \rhop \rhom$ as shown in Fig.~\ref{fig:rhorhocheck}a reveals that for the central values considered, the resulting plateau length of less than $0.01^\circ$ can be safely ignored. This begs the question on the point at which this effect should emerge. Clearly, this is dictated by the central value of $\BF^{00}/\BF^{+0}$, so a scan is performed by steadily increasing its value to a worst-case $\BF^{00}/\BF^{+0} = 0.1$ as can be seen in Fig.~\ref{fig:rhorhocheck}b. As long as the ratio remains roughly under $\BF^{00}/\BF^{+0} \sim 0.045$ beyond which the plateau visibly emerges, this method remains viable at \lhcb assuming $\BF^{+-}/\BF^{+0}$ stays more or less unchanged.

\begin{figure}[!htb]
    \centering
    \includegraphics[height=140pt]{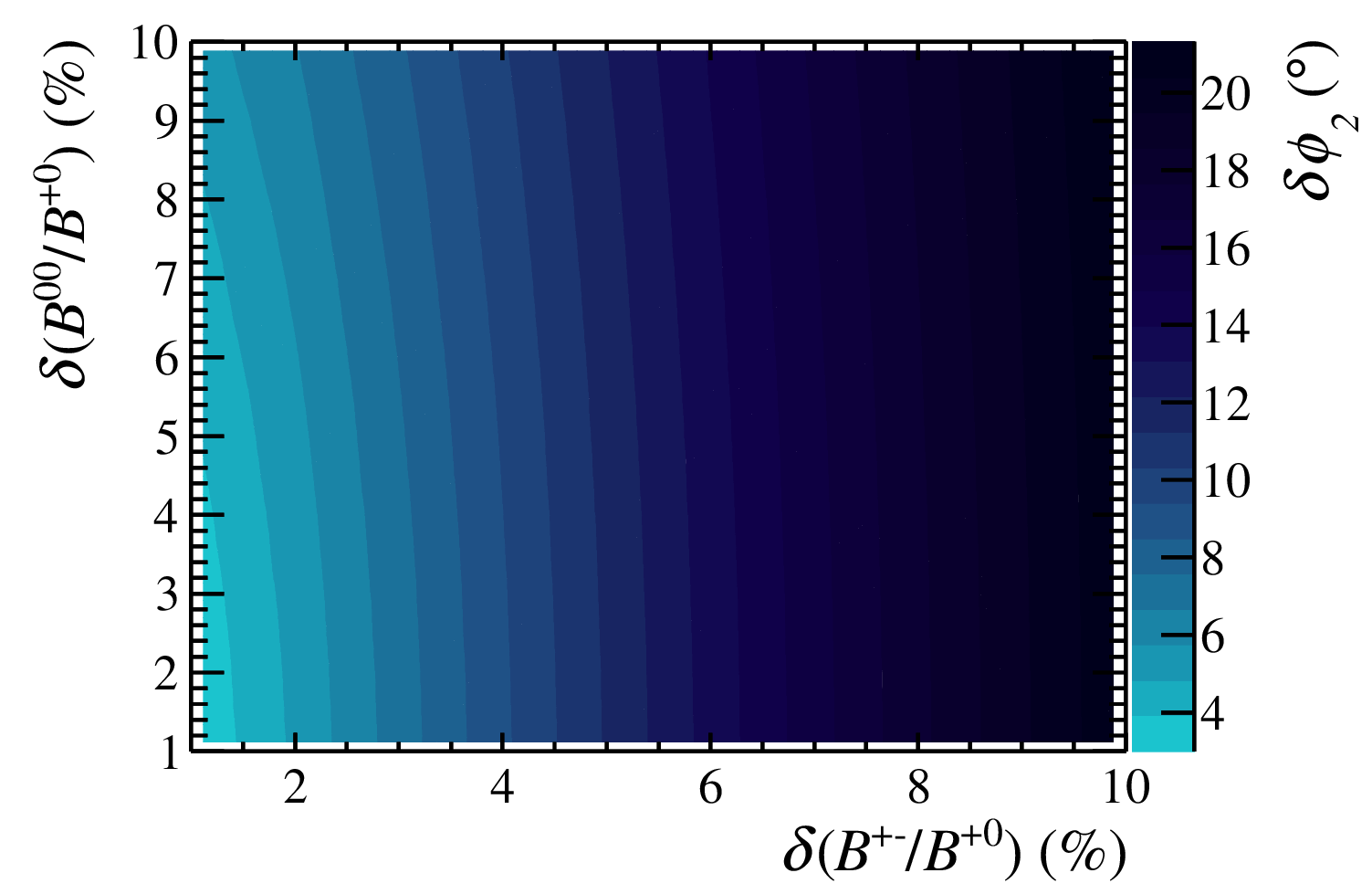}
    \put(-175,118){(a)}

    \includegraphics[height=140pt]{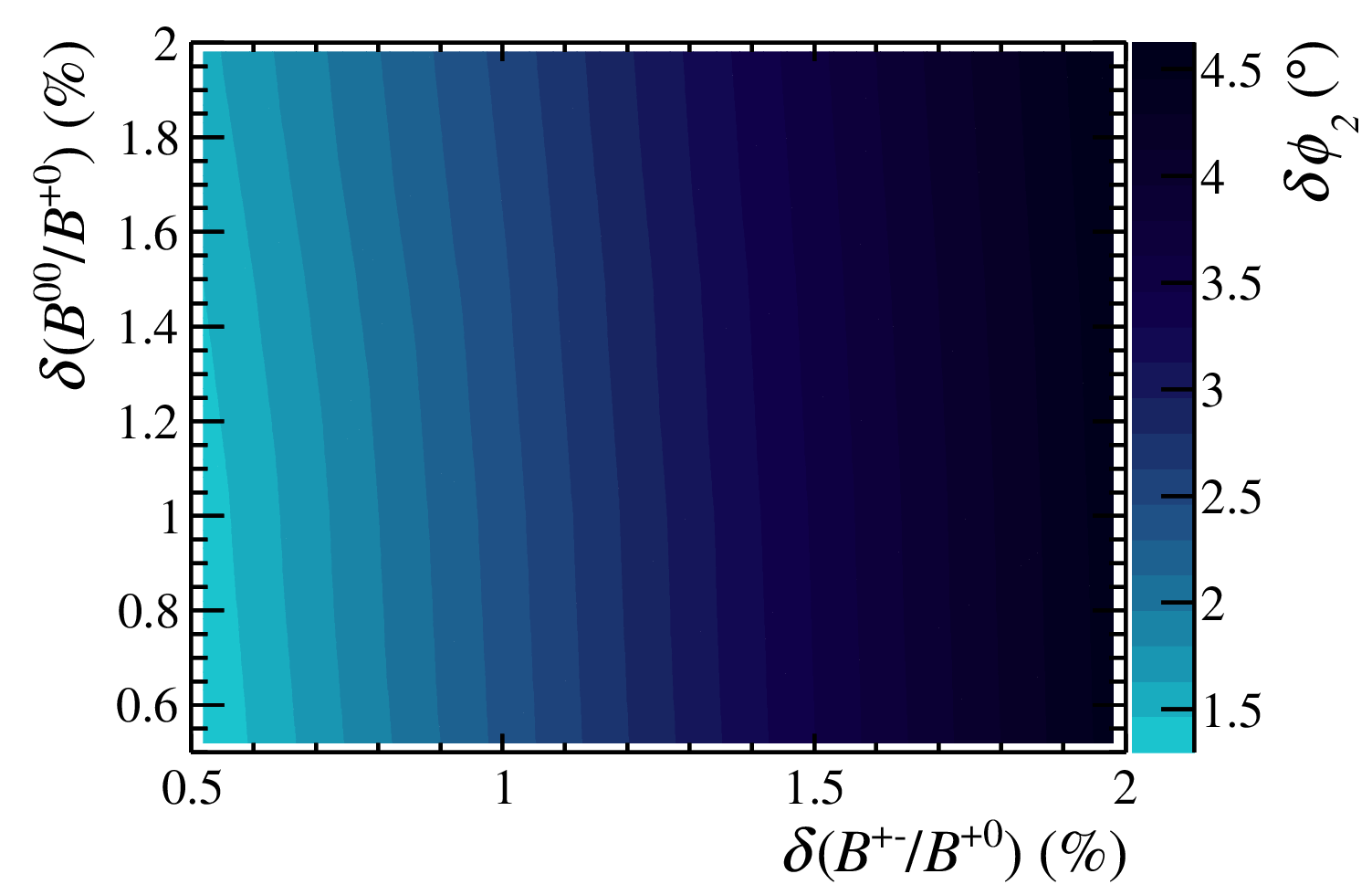}
%    \put(-393,118){(a)}
    \put(-175,118){(b)}

    \includegraphics[height=140pt]{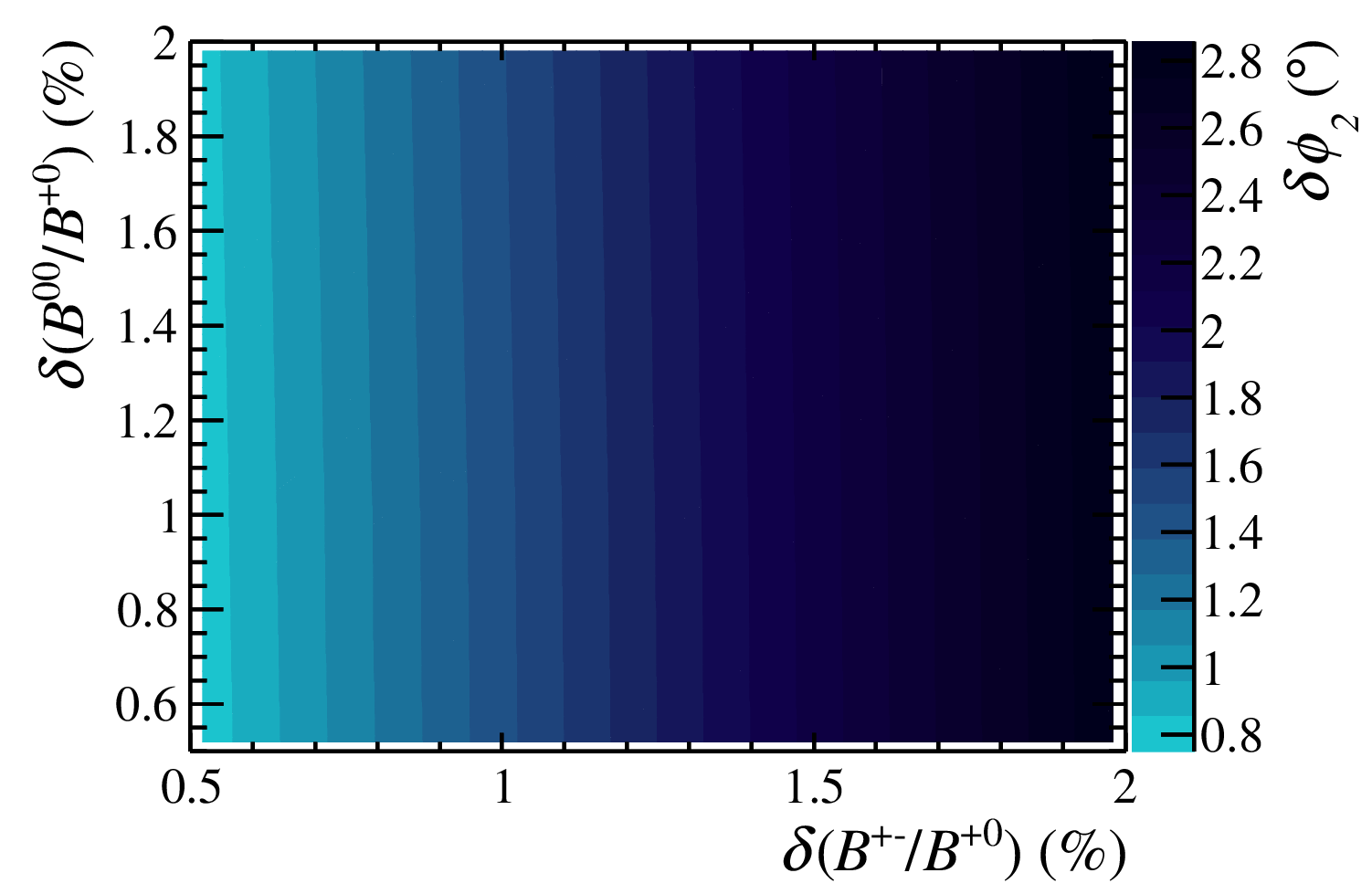}
    \put(-175,118){(c)}

    \caption{\label{fig:rhorhoscan} Projections of $\phi_2$ precision as a function of the total uncertainty in the $\BF^{+-}/\BF^{+0}$ and $\BF^{00}/\BF^{+0}$ relative branching fractions for (a)~Run~3, (b)~Run~5 and (c)~Run~5 with a time-dependent \CP-violation measurement in $\Bz \to \rhop \rhom$.}
\end{figure}

Three additional scans of the two-dimensional relative branching fraction uncertainty space are produced in Fig.~\ref{fig:rhorhoscan} showing the potential to constrain $\phi_2$ in Run~3, Run~5 and Run~5 again, assuming a time-dependent analysis of $\Bz \to \rhop \rhom$ becomes feasible. From the almost vertical bands in each scan, it is clear that the primary driver of $\phi_2$ precision will be the uncertainty in $\BF^{+-}/\BF^{+0}$, which makes sense as these sides of the isospin triangles are no longer the controlling lever arms. With an effective yield of only 500 events in $\Bz \to \rhop \rhom$, or in other words just 1\% of 1\% of the charmless four-charged-pion selection rate, the precision in this approach already looks competitive next to the first-generation \babar and \belle results. If a best-case 1\% total uncertainty can be attained by Run~3, the precision on $\phi_2$ will be roughly $3^\circ$. By Run~5, this will be driven down to $1.4^\circ$ and if a time-dependent analysis of $\Bz \to \rhop \rhom$ proves possible, an uncertainty of $0.8^\circ$ can be achieved, rivalling the \belletwo result.

A new source of systematic uncertainty in the ratio of branching fractions will be the \Bz to \Bp ratio of fragmentation fractions, $f_d/f_u$. To date, this has always been assumed to be unity due to isospin symmetry and thus strongly motivates its future study. It would also be prudent to survey the systematics landscape during Run~3, while there is still time to influence the design of the upgraded \lhcb electromagnetic calorimeter system~(ECAL) which is expected to be installed for the Run~5 data taking period and could prove critical in challenging the \belletwo legacy.

\section{Conclusion}
\label{sec:conclusion}

A rescaling of the SU(2) isospin triangles constraining $\phi_2$ that relies on measurements of the experimentally cleaner relative branching fractions is demonstrated to provide superior performance compared to the traditional approach relying on absolute branching ratios. Systematic uncertainties that cancel in these ratios lead to a markedly improved precision at \belletwo beyond current expectations, while the possibility to provide a competitive measurement in the $B \to \rho \rho$ system at \lhcb is enabled in consequence. An open question that remains is whether there is any significant reduction in isospin-breaking effects to be had within this approach.

\section*{Acknowledgements}
\begin{acknowledgement}
As always, I am greatly appreciative of the support from T.~Gershon, whose insight and careful reading improved this work immensely. I also wish to thank U.~Egede and J.~Zupan for useful exchanges on matters concerning systematic and theoretical uncertainties, respectively. Finally, I am indebted to my colleagues, B.~Adeva~Andany, V.~Chobanova and P.~Naik, for their assistance in providing helpful reviews for this paper. This work is supported by the ``Mar\'{i}a de Maeztu'' Units of Excellence program MDM2016-0692 and the Spanish Research State Agency. Financial support from the Xunta de Galicia (Centro singular de investigaci\'{o}n de Galicia accreditation 2019-2022) and the European Union (European Regional Development Fund – ERDF), is also gratefully acknowledged.
\end{acknowledgement}

%
% BibTeX users please use
\bibliographystyle{spphys}
\bibliography{main}

\end{document}